\documentclass[prl,showpacs,floatfix,amsmath,twocolumn,amsfonts]{revtex4}

\usepackage[dvips]{graphicx}

\usepackage{amsmath,amsthm}
\usepackage{epsfig}
\usepackage{amssymb}

\newcommand{\myIm}{\mathop{\mbox{Im}}}

\newcommand{\norm}[1]{\Vert #1 \Vert}

\begin{document}

\title{Collapse in boson-fermion mixtures with all-repulsive interactions}

\author{Vladyslav I. Prytula$^1$}
\author{Vladimir V. Konotop$^{1,2}$}
\author{V\'{\i}ctor M. P\'erez-Garc\'ia$^{1}$}
\author{Vadym E. Vekslerchik$^{1}$}
\affiliation{$^1$E.T.S. de Ingenieros Industriales, Departamento de Matematicas, and \\ Instituto de Matem\'atica Aplicada a la Ciencia y la Ingenier\'{\i}a, \\Universidad de Castilla-la Mancha, 
   Avda. Camilo Jose Cela, 3, 13071 Ciudad Real, Spain.
\\
$^2$Centro de F\'{\i}sica Te\'orica e Computacional,  Universidade de
Lisboa, Av. Prof. Gama Pinto 2, Lisboa 1649-003, Portugal and
Departamento de F\'{\i}sica, Faculdade de Ci\^encias,
Universidade de Lisboa, Campo Grande, Ed. C8, Piso 6, Lisboa
1749-016, Portugal.
}

\begin{abstract}
We describe  the collapse of the bosonic component in a boson-fermion mixture due to the pressure exerted on them by a large fermionic component, leading to collapse in a system with all-repulsive interactions. We describe the phenomena early collapse and of super-slow collapse of the mixture.
\end{abstract}

\maketitle

\emph{Introduction.-} The achievement of Bose-Einstein condensates (BECs) \cite{original} has pushed the field of degenerate quantum gases to become one of the most active areas of Physics. After the condensation of different bosonic species, degenerate mixtures of bosonic and fermionic atoms were created \cite{Truscott} 
providing a highly controllable tool for study of systems of mixed quantum statistics. 

Several interesting nonlinear phenomena have been studied in the context of boson-fermion mixtures. First, the fact that interspecies interactions may result in attraction among bosons~\cite{TW},
implies that solitons could also in quasi-one dimensional boson-fermion mixtures \cite{Karpiuk,skk,Solitons3}. Also, collapse of the atomic cloud induced by the interspecies attraction in boson-fermion mixtures was observed experimentally \cite{Collapse} and  studied theoretically \cite{theo1}. 

The usual scenario of collapse in single-species quantum gases corresponds to a BEC which collapses because of its attractive self-interaction 
\cite{Hulet2}.
This phenomenon had been long known in mathematical physics \cite{Sulem} but the theoretical description of  its specificities for BECs has motivated a lot of theoretical research.  In boson-fermion mixtures, it is the attractive interspecies interaction what drives 
 the blow-up of the atomic cloud \cite{Collapse}.

In this paper we describe new blow-up scenarios in degenerate quantum gases, namely in boson-fermion mixtures, when \emph{all interatomic interactions are repulsive}. More specifically we  will describe the phenomena of {\em early} collapse and of {\em super-slow} collapse of the mixtures.

\emph{The model.-} We describe bosons in the  mean field approximation and  spin polarized fermions at zero temperature in the hydrodynamic approximation, when the number of fermions is much larger than the number of bosons. The respective dynamical system reads~\cite{TW,BK}
\begin{subequations}\label{ZZZ}
\begin{eqnarray}
\label{GP}
i \hbar \frac{\partial \psi}{\partial t} =  - \frac{\hbar^2}{2m_b}\Delta \psi + V_b(\boldsymbol{r})\psi + g_{bb} |\psi|^2 \psi+ g_{bf}\rho\psi,
      \\ 
\label{wave}
\frac{\partial^2 \rho_1}{\partial t^2}   = 
    \nabla 
            \left[
                \rho_0 \nabla 
                    \left(
                       \frac{\left(6 \pi ^2\right)^{2/3}\hbar^2}{3m_f^2\rho_0^{1/3}}\rho_1+\frac{g_{bf}}{m_f}|\psi|^2
                    \right)
            \right], 
\end{eqnarray}
\end{subequations}
where $\psi(\boldsymbol{r},t)$ is the macroscopic wave function describing the bosonic component and $\rho_1(\boldsymbol{r},t)$ is the excitation of the density of fermions, due to boson-fermions interactions, from the unperturbed density $\rho_0(\boldsymbol{r})$, so that the total fermionic density is given by $\rho(\boldsymbol{r},t)= \rho_{0}(\boldsymbol{r}) + \rho_{1}(\boldsymbol{r},t)$. $g_{bb}$ and $g_{bf}$ are the coefficients of two-body boson-boson and boson-fermion interactions, $V_b(\boldsymbol{r})$ is the confining potential for bosons,  $V_b=m_b\omega_b^2r^2/2$.

It is convenient to introduce the scaled variables $\boldsymbol{x}=\boldsymbol{r}/a$ and $\tau = \omega_b t$, where $a=\sqrt{\hbar/m_b\omega_b}$ is the linear oscillator length for the bosons. We also define $n_0= (g_{bf}/\hbar \omega_b)\rho_0$, $n= (g_{bf}/\hbar \omega_b)\: \rho_1$ and $u = \sqrt{ g_{bf}m_b/\hbar \omega_b m_f}\psi$, and rewrite Eqs. (\ref{ZZZ}) in the form
\begin{subequations}
\label{Zak}
\begin{eqnarray}
\label{nls}
 i u_{\tau} & = & - \frac{1}{2}\nabla^2 u + V_0u + nu +  g|u|^2 u,
   \\ 
 n_{\tau\tau} & = & \nabla 
            \left[
                n_0 \nabla
                    \left(
                       \frac{\alpha}{n_0^{1/3}}n + |u|^2
                    \right)
            \right],
\label{wave1}
\end{eqnarray}
\end{subequations}
where $V_0\equiv V_0(x) = x^2/2+n_0(x)$, $g=g_{bb}m_f/g_{bf}m_b$,  $\alpha = \left(6\pi^2\right)^{2/3}\hbar^{5/3}m_b/3m_f^2\left(g_{bf}\right)^{2/3}\omega_b^{1/3}$, $\nabla$ is understood in terms of the new coordinates and $x \equiv |\boldsymbol{x}|$ for $\boldsymbol{x}\in \mathbb{R}^3$.
We will refer to this model as the {\em nonlinear Zakharov system} (NZS) since it can be seen as a generalization of the Zakharov system~\cite{Zakharov} describing Langmuir waves in plasmas, the latter being given by Eq. (\ref{Zak}) with $g_{bb}=0$ (i.e. in the absence of the the two-body interactions among bososns). We will consider Eqs. (\ref{Zak}) with 
 with analytical initial data $u(0,\boldsymbol{x})$, $n(0,\boldsymbol{x})$, and $n_\tau(0,\boldsymbol{x})$.
 
\emph{Conserved quantities.-} Equations (\ref{Zak}) can be rewritten in a Hamiltonian form. To this end, following \cite{Schochet_Weinstein}, we introduce the vector function
\begin{equation}
  \label{V}
  {\bf v} = -\int\limits_{0}^{\tau}n_0\nabla\left(\frac{\alpha}{n_0^{1/3}} n +|u|^2\right)d\tau,
  \end{equation}
 which describes the  hydrodynamic velocity of excitations and rewrite the NZS in the form
\begin{subequations}\label{Weinstein_Hamiltonian_Version}
\begin{eqnarray}
n_{\tau} & =&  - \nabla {\bf v},
\quad {\bf v}_{\tau}  =- n_0 \nabla\left(\frac{\alpha}{n_0^{1/3}} n+|u|^2\right),
    \label{eq:v}
    \end{eqnarray}
    \begin{multline}
   i u_{\tau}  = - \frac{1}{2} \nabla^2 u + \left[V_0+ \frac{n_0^{1/3}}{\alpha}
   \left(\frac{\alpha}{n_0^{1/3}} n+|u|^2\right)^2 
     \right.
\\ 
 \left.
     +g-\frac{n_0^{1/3}}{\alpha} |u|^2\right]u,
     \label{eq:u}
\end{multline}
\end{subequations}
which admits the Hamiltonian
\begin{multline}
  H = \int\left[ |\nabla u|^2 + \frac{1}{2}\left( g-\frac{n_0^{1/3}}{\alpha}\right)|u|^4 + V_0|u|^2
   \right.
   \\
   \left.
   + \frac{1}{2}\frac{n_0^{1/3}}{\alpha}\left(\frac{\alpha}{n_0^{1/3}}n+|u|^2 \right)^2 + \frac{n_0}{2}|{\bf v}|^2 \right] d\boldsymbol{x}
   \label{H}.
\end{multline}
(hereafter all integrals without explicit limits are taken over the whole space). $ H$ is an   integral of motion: $ H(\tau)\equiv  H(0)$.

Another integral of motion is given by the total number of bosons, $N_b=\|u\|_2^2 $, defined through the standard notation for the $L^p$ norm: $\|u\|_p=\left(\int |u|^pd\boldsymbol{x}\right)^{1/p}$ where $p$ is an integer. We also notice the $n$ describes excitations of the Fermi sea and thus the number of excited fermions is not conserved.

\emph{Modulational instability.-} The simplest manifestation of instabilities in nonlinear wave systems is the appearance of modulational instability. Neglecting the external trapping potential, i.e. supposing that $n_0$, and hence $g_0=n_0^{1/3}/\alpha$, are $\boldsymbol{x}$-independent constants, and looking for  plane-wave solutions of Eq. \eqref{nls} of the form $u_p= A \exp(i{\bf K}\boldsymbol{x} - i\Omega \tau)$ we get the corresponding excitation of the fermionic density $n_p = - g_0A^2$. Next we look for  perturbed solutions of the form  $u = u_p (1 +  u_1)$, 
$ n = n_p ( 1+   n_1)$, 
where $|u_1|\ll |u_p|$ and $|n_1|\ll |n_p|$. After linearizing with respect to $(u_1,n_1)\propto \exp(i{\bf k}\boldsymbol{x}-\omega t)$, from the resulting linear system we find the dispersion relation
\begin{eqnarray}
\label{disprel}
   2A^2g_0k^4 
   =
     ( \omega^2 - k^2 ) [ \omega^2 - 
      4({\bf k}\cdot{\bf K})^2 
     - 2gA^2k^2 - k^4 ]
\end{eqnarray}
The instability occurs when at least one of  the frequencies $\omega$ obtained from Eq. (\ref{disprel}) has a nonzero imaginary part. This happens for the wavevectors satisfying
\begin{equation}
    k^2 < 2A^2\left(g_0-g\right) - 4 K^2\cos^2\theta,
\end{equation}
where $\theta$ is the angle between ${\bf K}$ and ${\bf k}$. 
Thus the condition of modulational instability has the same functional form as that for the purely bosonic condensate \cite{Sulem} with the substitution of $-g$ by $g_0-g$. i.e. if $g_0>g$, then even when all two-body interactions are repulsive provided $g<g_0$ modulational instability may set in. From the physical point of view, the modulational instability develops for long wavelength excitations. For such excitations the fermionic excitations are smooth, and therefore the constant amplitude density $n_p$ enters the condition of the instability. Thus, this result means that instabilities may develop in a system with repulsive interatomic interactions, which is very different from the case of the purely bosonic condensate and is the first indication of the phenomenona to be discussed in detail in this paper.

\begin{figure}[h]
\epsfig{file=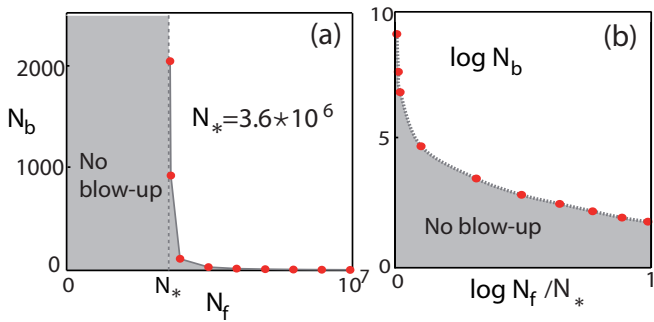,width=\columnwidth}
\caption{Ground state solutions of Eq. (\ref{ZZZ}) for $a_{bf} = 10\,$nm, $a_{bb} = 5.25\,$ nm, $m_f = 0.66 \times 10^{-25}$ kg, $m_b = 1.44 \times 10^{-25}$kg, $\omega_b = 188$ Hz, and $\omega_f = \omega_b \sqrt{m_b/m_f} =$ 269 Hz, corresponding to $^{87}$Rb-$^{40}K$ mixtures such as those of Ref. \cite{Collapse}.  Shown are (a) the location on the plane $N_b-N_f$ of the ground states satisfying   $H =4/27\gamma^2$, thus all ground state solutions in the shaded region must be stable. Panel (b) shows some more data on a logarithmic scale.} 
\label{una}
\end{figure}

\emph{Non-collapsing solutions.-} First, we will prove that there are initial data which do not blow,  i.e.  the existence of global solutions of Eq. (\ref{Weinstein_Hamiltonian_Version}). More specifically, we want to establish uniform boundness in time of $U=\|\nabla u\|_2^2$  \cite{Schochet_Weinstein,SS} for  $g<g_0$ (for $g>g_0$ the proof is trivial). Using the simplified version of  the Gagliardo-Nirenberg inequality $ \norm{w}_{4}^{4} \leq C_1 \norm{\nabla w}_{2}^{3} \norm{w}_{2}^{1}$ 
\cite{Cazenave}, with $C_1=0.44927\dots$ being the best constant in the 3D case~\cite{DelPino}, we get
\begin{align}
\label{neqH}
	H\geq  U^2-\gamma U^3, 
\end{align}
where $ \gamma = (g_0-g)C_1 N_b^{1/2}$. Eq. (\ref{neqH}) is valid for all $\tau$, and thus, if there is at least one positive root of  $F(U)\equiv \gamma U^3 -  U^2 + H= 0$, to be denoted as $u_+$: $F(u_+)=0$, and if
initially $U (0)<u_+,$ then Eq. (\ref{neqH}) guarantees that $U(\tau)<u_+$ for any time, i.e. $U$ is bounded uniformly in time. 
Hence, to establish the sufficient condition for the existence of the solution we have to find the condition of the existence of a positive root $u_+$. It is easy to see that $F(U)$ has a local minimum  at $U_{min} = 2/3\gamma$ and a sufficient condition for the existence of a positive root is 
   $ F(U_{min})\leq 0$, 
what is equivalent to the constraint
\begin{equation}
    H < H_0=\frac{4}{27}\frac{1}{\gamma^2}=\frac{4}{27}\frac{1}{(g_0-g)^2C^2_1N_b}\,.
    \label{1st_constraint}
\end{equation} 
It can be seen that the condition $U(0) < u_+$ is satisfied by choosing $U(0)^2\leq H$.   This rigorous result, being a sufficient condition, provides a lower bound for the largest energy allowing the existence of solutions, i.e. the physical region of existence could be larger, in particular allowing $H>H_0$. Since Eq. (\ref{1st_constraint}) involves both the Hamiltonian $H$ and the number of bosons $N_b$, which are not independent, in Fig. \ref{una} we plot the region of non collapsing ground states by fixing the number of fermions and computing numerically the ground state for different $N_b$ until the limit of the inequality is reached. This leads a domain in which we can guarantee the existence of global solutions. It is interesting to note that there exists a critical number of fermions, $N_*$ below which all solutions exist globally and no collapse can occur. The number of fermions separating the domain of globally existing solution decreases with $N_b$, what is explained by the fact that the repulsion between bosons due to two-body interactions  decreases with $N_b$, and thus smaller fermionic densities are able to dominate the inter-bosonic repulsion. 

\emph{Collapsing solutions.-} It was shown  numerically in \cite{Landman-Papanicolaou} that the Zakharov equations may undergo finite-time collapse. In our system, due to the presence of repulsive boson-boson interactions it may seem counter-intuitive that such a phenomenon could exist. However, we will argue that
\emph{the bosonic component in a Bose-Fermi mixture, in which there are many more fermions than bosons, can undergo collapse in finite time even when both the self-interaction between bosons and the interspecies interactions between bosons and fermions are repulsive}.
To support this statement we will use ``early collapse" type arguments \cite{Konotop_Paciani}, modified for the NZS since the standard arguments based on the virial identities cannot be used.

Let us define the squared wavepacket width as $Y(\tau) = \int x^2 |u|^2 \;d\boldsymbol{x}$. By direct derivation we obtain  $\dot Y(\tau)= 4\myIm\int \bar{u} \left( \boldsymbol{x} \cdot 
\nabla u \right)
d\boldsymbol{x}$ and
\begin{eqnarray}
\label{y_dot}
\ddot Y(\tau)= \int
\left(8|\nabla u|^2\;d\boldsymbol{x} + 6g
|u|^4
 -4 
 |u|^2 
 \boldsymbol{x}
 \nabla n  
 \right)d\boldsymbol{x},
 \label{Y''}
\end{eqnarray}
(hereafter an overdot stands for the derivative with respect to $\tau$). In our case 
and even in the case of classical Zakharov system  \cite{Merle_1996} the lack of control on the last term of Eq. (\ref{Y''})
makes impossible to apply the usual virial-type arguments. Nevertheless by analyzing Eq. (\ref{y_dot}) for small times we can extract information about the early stage of the evolution. First, we notice that $Y(0)$ and $\dot Y(0)$  depend only on the initial distribution of the boson component, $u(0,\boldsymbol{x})$. In particular, for simplification of the further analysis, we can make $\dot Y(0)=0$ by choosing $u(0,\boldsymbol{x})$ real. Considering the higher order derivatives one can note that for given $u(0,\boldsymbol{x})$,  $\ddot Y(0)$ can be made negative by an appropriate choice of $n(0,\boldsymbol{x})$. For example, this occurs if $n(0, \boldsymbol{x})$ is a cup-like profile. Moreover, from the expression for the third derivative %
\begin{eqnarray}
\label{Y'''}
\dddot Y(0) = -4 \int |u(0, \boldsymbol{x})|^2 \boldsymbol{x} \cdot \nabla n_{\tau}(0,{\boldsymbol{x}})\;d\boldsymbol{x},
\end{eqnarray}
it follows that $\dddot Y(0)$  can also be made negative by an appropriate choice of $n_{\tau}(0,x)$.

Summarizing the above arguments, assuming that $Y(\tau)$ is sufficiently smooth and requiring $\ddot Y(0)$ and $\dddot Y(0)$ to be negative, what is always possible by means of the proper choice of the initial conditions, we can expand $Y(\tau)$ in  Taylor series  for sufficiently small $\tau$: 
\begin{eqnarray}
Y(\tau)=(\tau^2-\tau_{*}^2)\frac{|\ddot Y (0)|}{2} + \frac{\tau^3}{6}\dddot Y(0)+ o(\tau^3).
\label{Taylor}
\end{eqnarray}
Here $\tau_*=\sqrt{-2Y(0)/\ddot Y(0)}$. Now we  notice that the higher terms that are not written explicitly in (\ref{Taylor})  depend on higher derivatives of the fields $u(\tau,\boldsymbol{x})$ and $n(\tau,\boldsymbol{x})$ at $\tau=0$ [this can be verified by direct computation of the higher derivative of $Y(\tau)$, by analogy with (\ref{Y''}) and (\ref{Y'''})], and thus can be chosen to satisfy {\it a priori} given properties (which in our case are reduced to existence of a nonzero radius of convergence of the Taylor series).  
This means that one can construct initial data such that $Y(\tau)$ crosses zero at $\tau\leq\tau_*$. This contradiction with the definition of $Y(\tau)$, which is nonnegative,  implies a finite time collapse for the chosen initial data. 

The above arguments, would become mathematically rigorous only if one could prove that $\tau_*$ is smaller than the convergence radius of the series (\ref{Taylor}). Although that proof is not available, we can show that by proper choice of the initial conditions one can make $\tau_*$ as small as necessary, what supports our conjecture. 

Let us consider a realistic situation, where initially the fermions are unperturbed and the bosons have a Gaussian distribution narrower than the size of the fermionic cloud.
In that situation, $n_0$ can be taken as constant and we can compute $\tau_*$, given below in the physical units:
\begin{equation}\label{T}
T_{*}=
 \frac{(2\pi)^{1/4}a^{1/2}}{\omega_b(N_ba_{bf})^{1/2}}\frac{m_f}{m_b+m_f} 
\left( \frac{(6\rho_0)^{1/3}a_{bf}}{\pi^{1/3}}   -\frac{2a_{bb}}{a_{bf}}\right)^{-1/2}.
\end{equation}
where $\omega_b$ is the initial width of the bosonic cloud. 
From Eq. (\ref{T})  it can be clearly seen that controlling $a_{bf}$ allows one to make $T_*$ as small as necessary. Taking the same data as in Fig.~\ref{una}, but with $a_{bf}=10^{-7}m$, we get a blow-up time $T_{*}\approx 0.127$ s.

\emph{Super-slow collapse.-} Finally, following  Ref. \cite{Merle_1996} where the possibility of collapse was studied for the conventional Zakharov system, we outline the proof of the fact that the bosonic component for $g<g_0$  and $H<0$ may undergo a completely different physical phenomenon: infinite time blow-up, which we will denote as \emph{super slow blow-up}. We restrict our analysis to radially symmetric solutions and for the sake of simplicity consider a spatially homogeneous mixture with $\alpha=1$ and $n_0=1$ (i.e. $g_0=1$), what can always be done by proper rescaling of the independent variables. Recalling that $x$ is a dimensionless radial variable we define
\begin{equation}
\label{def_z}
    Z(\tau)=\myIm\int_0^{\infty}
     \left(x^2 \bar{u} u_x  
        + x n {\bf v} \cdot {\bf x}  \right) f_x dx,
\end{equation}
where the weight function $f(x)$ is chosen, such that for arbitrary positive constants $R_0$, $C_2$ and $C_3$, the following conditions are satisfied~\cite{Ogawa_Tsutsumi}: 
\begin{equation}
\begin{array}{l}
     |f_x|\leq R_0, \quad f_{xx}<1,  \mbox{and}\quad  |\Delta^2_{r}f| \leq C_2, 
 \\  
     C_3\geq3-\Delta_{r}f\geq 0 \quad \mbox{for}\; x\geq R_0 
     \\
     3-\Delta_{r}f = 0 \quad \mbox{for}\; x <R_0,
\end{array}
     \label{app}     
\end{equation}
where $\Delta_{r} \equiv \frac{\partial^2}{\partial x^2}+\frac 2x \frac{\partial}{\partial x}$.
By the direct algebra we obtain
\begin{eqnarray*}
   -\dot Z(\tau)  = \int_0^{\infty} \left[ \frac{|u|^2}{2}\Delta^2_{r} f
+ (3-2f_{xx})|u_x|^2 
        + \frac{2}{x}f_x |{\bf v}|^2
        \right.
         \nonumber
        \\    
   + \left.
    (3-\Delta_{r}f)\left(
        n|u|^2  +\frac{g}{2}|u|^4 +\frac{n^2+|{\bf v}|^2}{2} 
     \right)\right]x^2dx
     - \frac{3H}{4\pi},
     \label{dotZ}
\end{eqnarray*}
 and unisng properties (\ref{app}) we estimate 
\begin{multline}
    -\dot Z(\tau)\geq
    \frac{g-1}{2}\int_0^{\infty} \left((3-\Delta_{r}f)
        |u|^4 + |u_x|^2\right)
    x^2dx     \\
     -C_2 -\frac{3H}{4\pi}.
     \label{dotZ2}
\end{multline}
Radially symmetric functions $u\in L^2(\mathbb{R}^3)$, with $|u_x|\in L^2(\mathbb{R}^3)$, satisfy the  Strauss lemma \cite{Strauss}: $|u|^2\leq \beta R^{-2}\norm{u}_{2}\norm{u_x}_{2}$ for $x>R$, where $R$ is an arbitrary positive constant and the constant $\beta$ is independent on $R$.
Since $g<1$, application of the Strauss lemma with $R=R_0$ together with (\ref{app}) enables us to estimate the right hand side of Eq. (\ref{dotZ2}) as
\begin{equation}
  - \dot Z \geq - \frac{3H}{4\pi}
  - \frac{C_2}{2}\norm{u}^2_2 
  - \frac{C_3 \beta^2 (1-g)^2}{16 R_0^4}\norm{u}^6_2.
\end{equation}
Thus if $H$ is negative, then choosing the constants $C_2$ and $R_0$ one can obtain
$
-\dot Z(\tau)>0.
$
On the other hand, following \cite{Merle_1996}, it can be shown that 
$
-Z(\tau) \leq C_4 \left[ 1 + \int_0^{\infty} \left( |u_x|^2 +n^2 +|{\bf v}|^2  \right)x^2dx\right] 
$ where $C_4$ is a positive constant depending only on $\norm{u}_2$ and on $|f_x|$. This leads to   
\begin{equation}
  \lim_{\tau \rightarrow \infty} \int_0^{\infty} \left( |u_x|^2 +n^2 +|{\bf v}|^2\right)\:x^2dx
  = \infty,
\end{equation}
which implies collapse for infinite times and physically gives an indication that a large fermionic component is able to slowly compress a  bosonic component even when two-body interactions of bosons are repulsing, provided they are not too strong. This leads to a \emph{super-slow} collapse. 

\emph{Conclusions.-} We have studied different collapse scenarios of the bosonic component in boson-fermion mixture at limited positive boson-boson  scattering lengths (defined by $g < g_0$) provided the fermionic component has the density large enough. They are (i) a finite-time collapse, occurring at early stages of the evolution, and (ii) super-slow collapse happening at infinite times. The physical reason for collapse in both cases is the pressure exerted on the bosons by the fermionic component. It is remarkable that these phenomena occur even when all interatomic interactions are repulsive.  Our theoretical predictions can be tested with current experimental setups and describe novel interesting singular quantum phenomena with degenerate quantum gases.

\acknowledgements

This work has been partially supported by grants  FIS2006-04190 and SAB2005-0195 (Ministerio de Educaci\'on y Ciencia, Spain), PAI-05-001 (Junta de Comunidades de Castilla-La Mancha, Spain)  and POCI/FIS/56237/2004 (FCT -Portugal- and European program FEDER).

\end{document}